%% file: CZgate_resubmit.tex
\documentclass[aps,groupedaddress,showpacs,letterpaper,twocolumn,reprint,nofootinbib,longbibliography]{revtex4-1}
\usepackage{graphicx} % standard LaTeX graphics tool
\usepackage{amsmath} 
\usepackage{amssymb}   % AMS tex symbols
\usepackage{color}  % for coloring edit command

\newcommand{\ket}[1]{| #1 \rangle}
\newcommand{\bra}[1]{\langle #1 |}

\begin{document}

\title{Symmetric Rydberg controlled-Z gates with  adiabatic pulses}
\author{M. Saffman}
\affiliation{
Department of Physics,
University of Wisconsin-Madison, 1150 University Avenue,  Madison, Wisconsin, 53706, USA
}
\author{I. I. Beterov}
\affiliation{Rzhanov Institute of Semiconductor Physics SB RAS, and 
Novosibirsk State University, 630090, Novosibirsk, Russia
}
\author{A. Dalal, E. J. P\'aez, and B. C. Sanders}
\affiliation{
Institute for Quantum Science and Technology, University of Calgary, Calgary, Alberta, T2N 1N4 Canada
}

 \date{\today}

\begin{abstract} We analyze  neutral atom Rydberg $C_Z$ gates based on adiabatic pulses applied symmetrically to both atoms.  Analysis with smooth pulse shapes and Cs atom parameters predicts the  gates can create Bell states with fidelity ${\mathcal F}>0.999$ using adiabatic rapid passage (ARP) pulses. With globally optimized adiabatic pulse shapes, in a two-photon excitation process, we generate Bell states with fidelity ${\mathcal F}=0.997$. The analysis fully accounts for spontaneous emission from intermediate and Rydberg states, including the Rydberg lifetime in a room temperature environment, but does not include errors arising from laser noise.
The gate protocols do not require individual addressing and are  shown to be robust against Doppler shifts due to atomic motion. 
\end{abstract}

\maketitle

\section{Introduction}

Qubits encoded in hyperfine states of neutral atoms can be entangled using $C_Z$ or CNOT gates mediated by Rydberg state 
interactions~\cite{Jaksch2000,Saffman2010}. Recent experimental progress has demonstrated entanglement fidelity of ${\mathcal F}\simeq0.97$ in a one-dimensional geometry~\cite{Levine2019} and ${\mathcal F}\simeq0.89$ in a two-dimensional qubit array~\cite{Graham2019}.  Numerous pulse protcols for  Rydberg entanglement have been proposed~\cite{Jaksch2000,Muller2009,Saffman2009b,Rao2014,Han2016,Su2016,Theis2016b,Beterov2016b,Petrosyan2017,Shi2017,Levine2019,Mitra2020}. Detailed analysis of gates using constant amplitude Rydberg excitation pulses has predicted a fidelity limit of ${\mathcal F}<0.999$ for Rb or Cs atoms in a room temperature background environment~\cite{XZhang2012}. Using Cs atoms the theoretical limit has been extended to  ${\mathcal F}> 0.9999$ with smooth, analytic pulses in~\cite{Theis2016b},
and with a dark state  mechanism in~\cite{Petrosyan2017}.

Although there now exist protocols for entangling gates that offer ${\mathcal F}> 0.9999$, a performance which is expected to be sufficient for scalable quantum 
computation~\cite{Devitt2013}, there is a  fidelity gap between predicted performance and the best experimental results~\cite{Levine2019,Graham2019}. Some of the missing fidelity can be ascribed to known technical imperfections including laser noise, Doppler broadening from finite atomic temperature, and the possible influence of background electric fields~\cite{Saffman2016,deLeseleuc2018}. 
Many Rydberg entanglement experiments have suffered from unexpectedly large loss of atoms that are left in Rydberg states during intermediate steps of the entanglement protocol~\cite{Isenhower2010,Zhang2010,Wilk2010,Maller2015,Jau2016,YZeng2017}. 

In order to reduce any excess Rydberg state loss it is desirable to develop protocols that do not leave population in Rydberg states where they are subject to relatively fast decoherence during intermediate stages of the gate. The standard Rydberg blockade $C_Z$ pulse sequence~\cite{Jaksch2000} consists of a $\pi$ pulse on the control qubit, a $2\pi$ pulse on the target qubit, and a $\pi$ pulse on the control qubit, with each pulse resonant between a ground hyperfine qubit state $\ket{1}$ and a Rydberg level.  If the control qubit enters the gate in state $\ket{1}$ it is Rydberg excited and will  sit in the Rydberg level during the $2\pi$ pulse on the target qubit. It has been observed that this leads to larger loss, as compared to the case of a continuous $2\pi$ pulse on the ground - Rydberg transition~\cite{Maller2015,Graham2019}. For this reason it appears advantageous to develop protocols that continuously drive the ground - Rydberg - ground transition. We note that since the logical action of a $C_Z$ gate in a quantum circuit is symmetric with respect to the control and target inputs, it is natural to seek a gate protocol that is also symmetric with respect to interchange of qubits as regards the applied pulses.  

There has been previous work using symmetric  driving of both atoms for Rydberg gate protocols.  
Entanglement was demonstrated using continuous driving of a ground-Rydberg-ground transition on both atoms with constant 
amplitude pulses~\cite{Wilk2010}, and with symmetric but not continuous pulses in~\cite{Levine2019}. A $C_Z$ gate protocol with continuous driving of both atoms is possible using an adiabatic pulse with time varying Rabi frequency $\Omega(t)$ and Rydberg level detuning $\Delta(t)$ in the limit of $|\Omega|>{\sf B}$, where $\sf B$ is the Rydberg-Rydberg interaction strength~\cite{Jaksch2000}. Unfortunately, the requirement of adiabaticity renders the gate slow and susceptible to spontaneous emission from the Rydberg level. A careful optimization of the pulse parameters resulted in prediction of a 
fidelity of not more than ${\mathcal F}\sim 0.98$~\cite{Muller2014}. A version of the adiabatic gate using adiabatic rapid passage (ARP) pulses has yielded ${\mathcal F}=0.995$\cite{Mitra2020} and  together with electric field switching of Rydberg F\"orster resonances ${\mathcal F}=0.996$\cite{Beterov2016b}. 
A related proposal for an adiabatic gate in the intermediate regime $|\Omega|\sim \sf B$  yielded ${\mathcal F}\sim 0.95$~\cite{Rao2014}. Analysis of  entanglement creation with ${\mathcal F}< 0.999$ based on stimulated rapid adiabatic  passage (STIRAP) pulses and evolution of a two-atom dark state in the blockade regime of $|\Omega|\ll \sf B$ was presented in~\cite{Moller2008}. It was also shown how to implement an adiabatic phase gate, but a complete fidelity analysis was not performed.  Additional variations of adiabatic pulses for entangling gates were analyzed in~\cite{Goerz2014} although the effect of a finite Rydberg lifetime was not included, and in \cite{Tian2015,HWu2017}.  

It is also possible to achieve a fast phase gate using non-adiabatic, constant amplitude pulses that continuously drive   both atoms. The challenge in making this work is that when only one atom is in the ground state that is Rydberg coupled (input states $\ket{01}$ or $\ket{10}$) the coupling rate is given by the one-atom Rabi rate $\Omega.$  However, if the input state is $\ket{11}$ the effective Rabi rate due to the Rydberg blockade mechanism is $\Omega_2=\sqrt2 \Omega$. Therefore a $2\pi$ pulse for states $\ket{01}$ or $\ket{10}$ will be a $2^{3/2}\pi$ pulse for $\ket{11}$ leading to large gate errors. 
This problem was solved in~\cite{Han2016,Su2016} using a combination of a larger pulse area and finite detuning $\Delta$ from the Rydberg level. Unfortunately, due to the larger pulse area, there is increased spontaneous emission from the Rydberg state and predicted gate fidelities are less than  ${\mathcal F}\sim 0.999$. 

In this paper we revisit adiabatic $C_Z$ protocols in the blockade regime of $|\Omega|\ll \sf B$,  and show that using rapid adiabatic methods~\cite{Bergmann1998}, together with optimized pulse shapes, we can achieve $C_Z$ gates with high fidelities. With ARP pulses that drive the ground-Rydberg transition of both atoms simultaneously,  and are almost continuous, we reach  ${\mathcal F}>0.999$. With STIRAP pulses the fidelity is  ${\mathcal F}\simeq 0.98-0.99$ for analytical pulse shapes, which we further improve to ${\mathcal F}=0.997$ with  optimized pulses that we refer to as a STIRAP inspired gate. The design methodology for either ARP or STIRAP
type versions of the gate is closely related to adiabatic protocols that have been previously studied for gates acting on multi-atom ensemble qubits~\cite{Beterov2013a,Beterov2016}. 

The rest of the paper is organized as follows. In Sec. \ref{sec.ARP} we present ARP pulses for implementing a $C_Z$ gate, calculate the resulting Bell state fidelity, and demonstrate the gate has improved robustness with respect to laser detuning or intensity variations compared to the standard gate with constant pulses which is analyzed in the appendix. In Sec. \ref{sec.STIRAP} we analyze a $C_Z$ gate based on symmetric driving with STIRAP pulses. Two versions of analytical pulse shapes are presented in Sec. \ref{sec.STIRAP1}, and in Sec. \ref{sec.STIRAP2} we consider STIRAP-inspired globally optimized pulses that provide higher fidelity entanglement. The results are summarized in Sec. \ref{sec.discussion}. 

 %%%%%%%%%%%%%%%%FIGURE%%%%%%%%%%%%%%%%
\begin{figure}[!t]
\includegraphics[width=7.7cm]{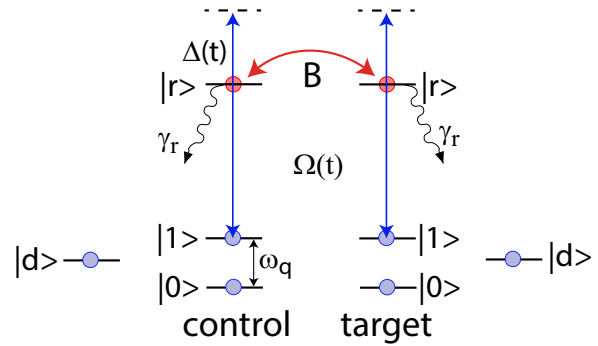}
\caption{(color online) Energy level structure of neutral atom qubits with ground states $\ket{0},\ket{1}$. Rydberg states $\ket{r}$ interact with strength $B$. One-photon excitation with Rabi frequency $\Omega(t)$ at detuning $\Delta(t)$. Level $\ket{d}$ is an uncoupled state that accumulates spontaneous emission from $\ket{r}$ which has lifetime $\tau_r=1/\gamma_r$ and decays to states $\ket{0},\ket{1},\ket{d}$ with branching ratios $b_{0r}=1/16,b_{1r}=1/16,b_{dr}=7/8$.
   }
\label{fig.states1}
\end{figure}
%%%%%%%%%%%%%%%%%%%%%%%%%%%%%%%%

\section{$C_Z$ gate with ARP pulses}
\label{sec.ARP}

\begin{figure}[!t]
\includegraphics[width=7.7cm]{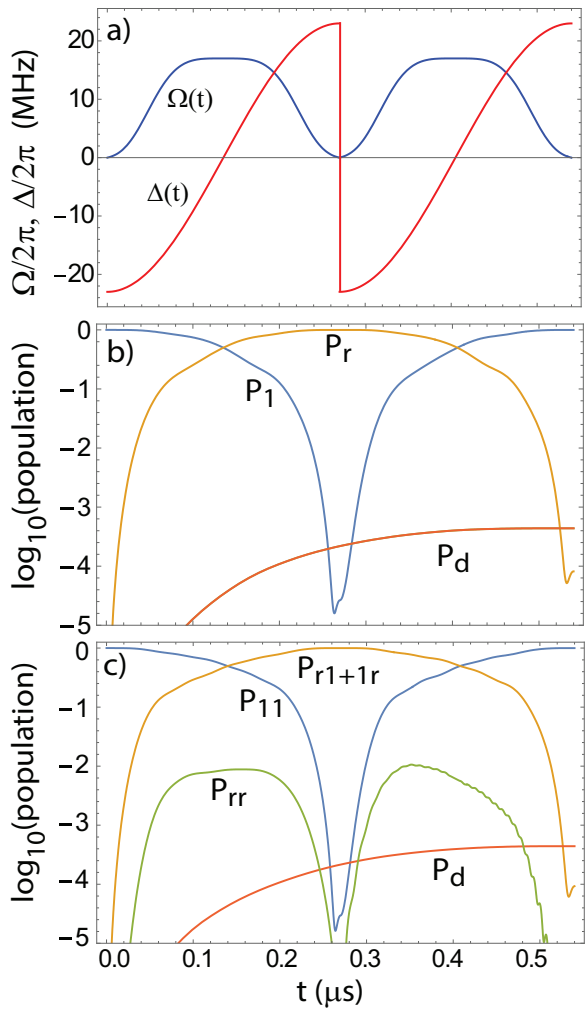}

\caption{(color online) $C_Z$ gate with ARP pulses.~a)~Time dependence of $\Omega(t)$ and $\Delta(t)$. b) Populations of the $\ket{10}$, $\ket{r0}$ and $\ket{d}$ states for the initial state  $\ket{10}$. The population in $\ket{d}$ is defined as $1-{\rm Tr}_{0,1,r}[\rho]$.  
c) Populations of the $\ket{11}$, $\ket{1r}+\ket{r1}$, $\ket{rr}$ and $\ket{d}$ states for the initial state  $\ket{11}$.
Parameters were $\Omega_{\rm max }/2\pi= 17 ~\rm MHz,$
$\Delta_{\rm max}/2\pi = 23 ~\rm MHz$, $B/2\pi = 100~\rm MHz$, total gate time $T=0.54 ~\mu\rm s$, $\gamma_r=1/(540 ~\mu{\rm s})$ 
$b_{dr}=7/8$, $b_{0r}=b_{1r}=1/16$.  The atomic parameters correspond to the Cs $107p_{3/2}$ state with spontaneous decay randomly distributed among the 16 ground hyperfine states. The temporal shape of the detuning was a quarter period of a $\sin$ function for each pulse and the Rabi drive was of the form $\Omega(t)=\Omega_{\rm max} \left[e^{-(t-t_0)^4/\tau^4}-a\right]/(1-a) $ with $t_0$ the center of the pulse, and the offset $a$ set to give zero amplitude at the start and stop points. For the data in the figure each pulse had a duration of $T/2$ and $\tau=0.175 T$.}
\label{fig.arp1}
\end{figure}

Consider excitation of a ground state $\ket{1}$ to Rydberg state $\ket{r}$ with  a one-photon transition  as shown in Fig.~\ref{fig.states1}. States $\ket{1},\ket{r}$ are coupled by a laser giving Rabi frequency $\Omega(t)$ and detuning $\Delta(t)$. The $C_Z$ protocol relies on driving a $2\pi$ rotation on both atoms.  The asymmetric states evolve as $\ket{01}\rightarrow  \ket{0r}\rightarrow e^{\imath\phi_1}\ket{01}$, $\ket{10}\rightarrow  \ket{r0}\rightarrow e^{\imath\phi_1}\ket{10}$, whereas the symmetric state $\ket{00}$ is dark to the gate pulses and in the limit of strong blockade 
the state $\ket{11}$ evolves as $\ket{11}\rightarrow  \frac{\ket{r1}+\ket{1r}}{\sqrt2}\rightarrow e^{\imath\phi_2}\ket{11}$.
The logical transformation in the basis $\{\ket{00},\ket{01},\ket{10},\ket{11} \}$ is therefore ${\rm diag}[1,e^{\imath\phi_1},e^{\imath\phi_1},e^{\imath\phi_2}]$. To achieve  a maximally entangling  $C_Z$ gate we may set $\phi_1=\phi_2=\pi$. This can be achieved by using a double ARP pulse with the detuning  reversed in the second pulse\cite{Beterov2013a} as shown in Fig.~\ref{fig.arp1}a).

%%%%%%%%%%%%%%%%FIGURE%%%%%%%%%%%%%%%%

The gate was analyzed  by numerical integration of the two-atom master equation in Lindblad form
\begin{equation}
\frac{d\rho}{dt}=i[H,\rho]+{\mathcal L}[\rho]
\label{eq.master}
\end{equation}
 with initial conditions
$\rho(0)=\rho_{\rm c}(0)\otimes\rho_{\rm t}(0)$ where $\rm c,t$ label control and target qubits.  
For the Hamiltonian we use 
${\mathcal H}={\mathcal H}_{\rm c}\otimes I + I\otimes {\mathcal H}_{\rm t} + {\sf B}\ket{rr}\bra{rr}$
with 
$$
{\mathcal H}_{c/t}=\left[\frac{\Omega(t)}{2}\ket{r}_{c/t}\bra{1}+ {\rm H.c.}\right]  +\Delta(t)\ket{r}_{c/t}\bra{r} ~.
$$
We neglect optical excitation of the $\ket{0}$ state due to the large detuning, $\omega_q\gg |\Omega|.$ We verify below that the additional gate errors from off-resonant excitation of this state, as well as off-resonant excitation of neighboring Rydberg states, is negligible for the chosen parameters.  
The decay term is  
$$
{\mathcal L}[\rho]=\sum_{\ell=c,t}\sum_{j=0,1,d} {L_j^{(\ell)}}\rho {L_j^{(\ell)}}^\dagger - \frac12{L_j^{(\ell)}}^\dagger {L_j^{(\ell)}}\rho -\frac12\rho {L_j^{(\ell)}}^\dagger {L_j^{(\ell)}}  
$$
with ${L_j^{(\ell)}}=\sqrt{b_{jr}\gamma_r}\ket{j}_\ell\bra{r}$ where $\gamma_r=1/\tau_r$ is the population decay rate of the Rydberg state and the $b_{jr}$ are branching ratios to lower level $j$. 
 The  levels $\ket{0},\ket{1},\ket{d}$ are taken to be stable. The uncoupled state $\ket{d}$  represents all the ground hyperfine states outside the qubit basis. Although some of these states are at the same energy as $\ket{1}$, and can be 
resonantly excited, we neglect such dynamics, thereby making the worst case assumption that all population leakage into $\ket{d}$ is an uncorrectable error.  

%%%%%%%%%%%%%%%%FIGURE%%%%%%%%%%%%%%%%
\begin{figure}[!t]
\includegraphics[width=8.cm]{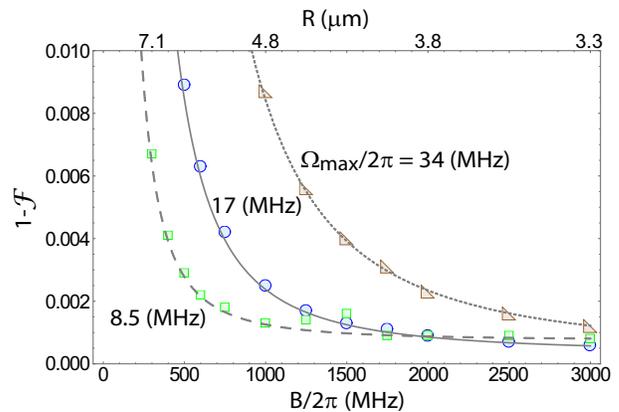}
\caption{(color online) Bell state infidelity using $C_Z$ gate with ARP pulses as a function of the interaction strength $\sf B$. Squares, circles, triangles  show results for $\Omega_{\rm max}/2\pi = 8.5, 17, 34 ~\rm MHz$, $\Delta_{\rm max}/2\pi = 11.5, 23, 46 ~\rm MHz$, and $T=1.08, 0.54, 0.27~\mu\rm s$, respectively. The lines are fits to
$1-{\mathcal F}= b + 7 (\Omega_{\rm max}/{\sf B})^2$ with $b=(7.5, 3.5, 3.2)\times 10^{-4}$. The upper abscissa axis shows the corresponding interatomic distance for  Cs $107p_{3/2}, m=3/2$ states with the quantization axis at 90 deg. to the line joining the atoms.  } 
\label{fig.arp2}
\end{figure}

Figure \ref{fig.arp1} shows the population evolution for states $\ket{10}$ and $\ket{r0}$ for parameters corresponding to excitation of the Cs $107p_{3/2}$ state. For both one and two Rydberg coupled atoms the populations faithfully execute a $2\pi$ rotation between ground and singly excited Rydberg states.  Although we are analyzing the ARP protocol as a one-photon excitation process, it could also be implemented as a two-photon transition for each pulse. This can be done for example by keeping the frequency of the first photon constant, with nonzero detuning from an intermediate level, and sweeping the frequency of the second photon. The one-photon analysis gives an upper limit to the gate fidelity. A two-photon implementation will suffer additional errors due to scattering from the intermediate state. The additional error can be made negligible provided there is sufficient laser power available to allow for large intermediate state detuning. It is also the case that for both one- and two-photon implementations there will be additional contributions to $\Delta(t)$ from the dynamic Stark shifts of the ground and Rydberg states. We do not explicitly include these shifts in the analysis. They can be corrected for either by modifying $\Delta(t)$ to compensate the Stark shifts, or by adding frequency sidebands to the excitation lasers to cancel the shifts. 

To generate entanglement we start with the state $\ket{ct}=\ket{11}$, apply a Hadamard gate to each qubit, the Rydberg $C_Z$ operation, and a final Hadamard to the target qubit which  ideally prepares the Bell state $\ket{B}=\frac{\ket{00}+\ket{11}}{\sqrt2}.$ The Bell fidelity can then be defined as~\cite{Sackett2000} ${\mathcal F} = \frac{\rho_{0000}+\rho_{1111}}{2}+|\rho_{1010}|.$ Assuming perfect Hadamard operations, and the same parameters as in Fig.~\ref{fig.arp1}  we find a Bell fidelity of ${\mathcal F = 0.9994}$ at ${\sf B}/2\pi=3~\rm GHz$ which is close to the maximum possible for the Cs 
$107\mathrm{p}_{3/2}$ state using the pulse shapes from Ref.~\cite{Theis2016b}. The fidelity exceeds 0.99 for ${\sf B}/2\pi=300~\rm MHz$ with the dependence of fidelity on $\sf B$ or, equivalently, interatomic spacing $R$  shown in Fig.~\ref{fig.arp2}.  The infidelity is accurately described by a small offset due to spontaneous emission, plus the scaling  $(\Omega_{\rm max}/{\sf B})^2$ which reflects blockade leakage allowing for finite excitation of the $\ket{rr}$ state.

The highest fidelity  result  uses a maximum Rabi frequency of only 17 MHz which implies leakage errors due to excitation of  $\ket{0}$, or due to excitation of $\ket{1}$ to a different Rydberg level that are bounded by $\epsilon\sim (\Omega_{\rm max}/\Delta_{\rm min})^2$ where $\Delta_{\rm min}$ is the smallest of $\omega_q$ or any of the detunings from $\ket{0}$ or $\ket{1}$ to nearby Rydberg levels. As detailed in~\cite{Theis2016b} for the Cs $107p_{3/2}$ state $\Delta_{\rm min}/2\pi \simeq 3 ~\rm GHz$ giving $\epsilon= 3\times 10^{-5}$ which is negligible relative to the calculated fidelity. 

\begin{figure}[!t]
    \centering
    \includegraphics[width=7.7cm]{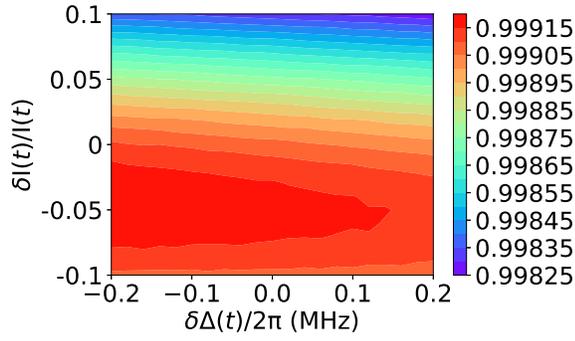}
    \caption{(color online) Robustness plot for the ARP pulses of Fig.~\ref{fig.arp1} with ${\sf B}/2\pi = 2.5~\rm GHz$. The plot shows  
     $\mathcal{F}$ with respect to changes in the detuning ($\delta\Delta(t)$) and fractional changes in the  intensity ($\delta \text{I}(t)/\text{I}(t)$)  of the laser. The highest fidelity is not at the center of the plot, indicating that slightly non-optimal parameters were used in Fig. \ref{fig.arp1}.}
    \label{fig.arp_robustness}
\end{figure}

Beyond the convenience of symmetric driving of the qubits, the use of adiabatic pulses makes the gate less sensitive to Doppler detuning and less sensitive to small variations in laser amplitude than a gate using constant amplitude pulses. As shown  in Fig.~\ref{fig.arp_robustness}  variation of the laser intensity and detuning of the ARP  gate  by $\pm 5~\%$ and $\pm 200~\rm kHz$ respectively reduces  the Bell fidelity by less than $0.0002$ so it stays above $0.999$. The allowance of $\pm 200~\rm kHz$ Doppler detuning is larger than the $\pm 80~\rm kHz$ Doppler shift variation for a Cs atom cooled to $10~\mu\rm K$ and Rydberg excited via a one-photon transition.  This insensitivity to parameters can be compared with the performance of the standard protocol with constant amplitude pulses presented in the Appendix Fig. \ref{fig.appendix}a). For the non-adiabatic gate of the same total duration, assuming the same intensity and detuning variations, the fidelity is reduced by more than $0.005$ which is  more than 10 times higher sensitivity to parameter values.  

\section{$C_Z$ gate with STIRAP pulses}
\label{sec.STIRAP}

\begin{figure}[!t]
\includegraphics[width=7.7cm]{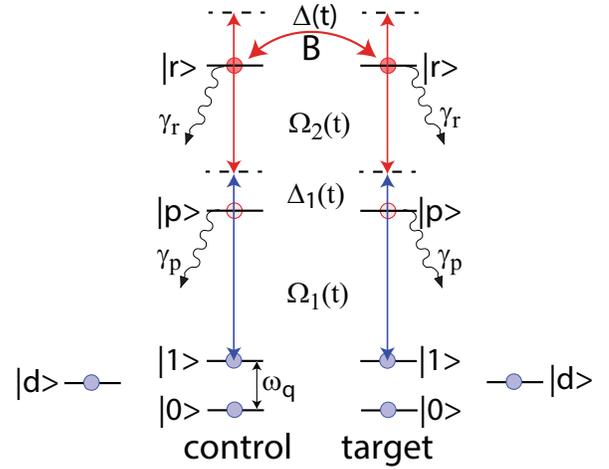}
\caption{(color online)  Two-photon excitation with Rabi frequencies $\Omega_1(t), \Omega_2(t)$ proceeds via intermediate state $\ket{p}$ 
at detuning $\Delta_{1}$. Level $\ket{d}$ is an uncoupled state that accumulates spontaneous emission from $\ket{p},\ket{r}$. State $\ket{p}$ has lifetime $\tau_p=1/\gamma_p$ and decays to states $\ket{0},\ket{1},\ket{d}$ with branching ratios 
$b_{0p},b_{1p},b_{dp}$. State $\ket{r}$ has lifetime $\tau_r=1/\gamma_r$ and decays to states $\ket{0},\ket{1},\ket{d},\ket{p}$ with branching ratios $b_{0r},b_{1r},b_{dr},b_{pr}$. 
   }
\label{fig.states}
\end{figure}

A STIRAP version of the gate using two-photon excitation is also possible as originally suggested in~\cite{Moller2008}.
Consider excitation of a ground state $\ket{1}$ to Rydberg state $\ket{r}$ with  two photons that are near resonant with intermediate state $\ket{p}$ as shown in Fig.~\ref{fig.states}. States $\ket{1},\ket{p}$ are coupled by a laser giving Rabi frequency $\Omega_1(t)$ and detuning $\Delta_{1}(t)$, while states $\ket{p},\ket{r}$ are coupled by a second laser with Rabi frequency $\Omega_2(t)$ and  detuning $\Delta_2(t)$. The two-photon detuning is $\Delta(t)=\Delta_{1}(t)+\Delta_{2}(t)$ which will be set to zero for resonant excitation.
The gate is again modeled by Eq.~(\ref{eq.master}) with the replacements 
\begin{eqnarray}
    {\mathcal H}_{\rm c/t}
    &=&\left[\frac{\Omega_1(t)}{2}\ket{p}_{\rm c/t}\bra{1}+\frac{\Omega_2(t)}{2}\ket{r}_{\rm c/t}\bra{p}+ {\rm H.c.}\right]\nonumber\\
&+&\Delta_1(t)\ket{p}_{\rm c/t}\bra{p} +\Delta(t)\ket{r}_{\rm c/t}\bra{r} 
\label{eq.H2}
\end{eqnarray}
and 
$$
{\mathcal L}[\rho]=\sum_{\ell=c,t}\sum_{j,k=0,1,d,p,r} L_{jk}^{(\ell)}\rho {L_{jk}^{(\ell)}}^\dagger - \frac12{L_{jk}^{(\ell)}}^\dagger {L_{jk}^{(\ell)}}\rho -\frac12\rho {L_{jk}^{(\ell)}}^\dagger {L_{jk}^{(\ell)}}   
$$
where  ${L_{jk}^{(\ell)}}=\sqrt{b_{jk}\gamma_k}\ket{j}_\ell\bra{k}$ for $j<k$ and $0$ otherwise. As with the analysis of the one-photon excitation ARP protocol we neglect any re-excitation of atoms that decay to $\ket{d}$. 

 %%%%%%%%%%%%%%%%FIGURE%%%%%%%%%%%%%%%%
%%%%%%%%%%%%%%%%%%%%%%%%%%%%%%%%

\subsection{Analytical STIRAP pulses}
\label{sec.STIRAP1}

%%%%%%%%%%%%%%%%FIGURE%%%%%%%%%%%%%%%%
\begin{figure}[!t]
\includegraphics[width=8.cm]{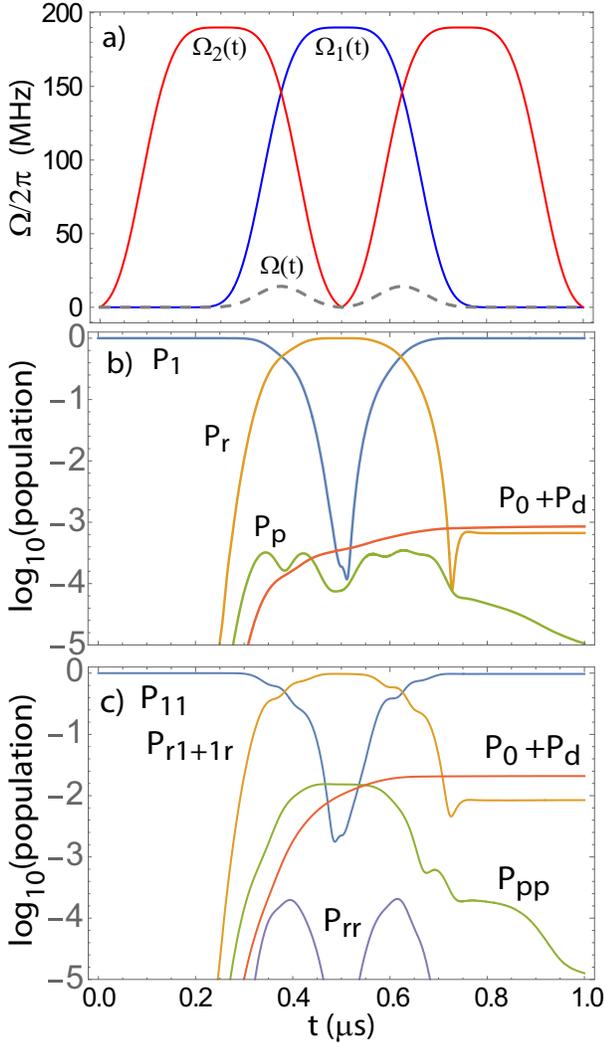}
\caption{(color online)
$C_Z$ gate with STIRAP pulses and total time of $T=1~\mu\rm s$. a) Pulse shapes.  The dashed gray line is the two-photon Rabi frequency $\Omega(t)=\Omega_1(t)\Omega_2(t)/2\Delta_1$. 
b) Populations of the $\ket{10}$, $\ket{p0}$, $\ket{r0}$,  and $\ket{00}+\ket{d0}$ states for the initial state  $\ket{10}$.  
c) Populations of the $\ket{11}$, $\ket{pp}$, $\ket{1r}+\ket{r1}$, $\ket{rr}$ and $\ket{0}+\ket{d}$ states for the initial state  $\ket{11}$. The curve labeled $P_{\rm pp}$ shows the total population in $\ket{p}$  which is defined as 
$2 \rho_{\rm pppp}+\sum_{j=0,1,\rm d,r}(\rho_{jj\rm pp}+\rho_{{\rm pp}jj}).$
Parameters were $\Omega_{1, \rm max }/2\pi= \Omega_{2, \rm max }/2\pi = 190~\rm MHz,$
$\Delta_1/2\pi = 750 ~\rm MHz$, $\Delta=0$, ${\sf B}/2\pi = 500~\rm MHz$, $\tau_p= 0.155~\mu\rm s$, $\tau_r=540.~\mu\rm s$\cite{2019stirapnote}, 
$b_{dp}=7/8$, $b_{0p}=1/16$, $b_{1p}=1/16$, 
$b_{dr}=7/16$, $b_{0r}=1/32$, $b_{1r}=1/32$, $b_{pr}=1/2$. 
 The Rabi pulses were of the form $\Omega(t)=\Omega_{\rm max} \left[e^{-(t-t_0)^4/\tau^4}-a\right]/(1-a) $ with $t_0$ the center of the pulse, and the offset $a$ set to give zero amplitude at the start and stop points. For $\Omega_1$ the pulse was centered at $t=T/2$ with $\tau=0.165 T$. For $\Omega_2$ the pulses were centered at $T/4, 3T/4$ with $\tau=0.175 T$.
}
\label{fig.stirap1}
\end{figure}

The requirement for a phase gate  is that the pulse shapes $\Omega_1(t), \Omega_2(t)$, and detunings $\Delta_1, \Delta$, are chosen such that all three states 
$\ket{01},\ket{10},\ket{11}$ return to the ground state after the applied pulse with phases $\phi_1, \phi_1, \phi_2.$   This is possible using the counterintuitive STIRAP sequence with $\Omega_2(t)$ preceding $\Omega_1(t)$, as shown in Fig.~ \ref{fig.stirap1}a). The population dynamics for one and two atoms in the Rydberg coupled state show high accuracy transfer to the Rydberg state and back to the ground state, as is shown in panels b) and c). Using the STIRAP pulse sequence, states $\ket{01}$ and $\ket{10}$ follow an adiabatic dark state with zero eigenvalue so there is no dynamical phase accumulation and $\phi_1=0$. An entangling $C_Z$ gate can then be obtained if $\phi_2=\pi$. If the intermediate STIRAP detuning is set to $\Delta_1=0$ the state $\ket{11}$ will follow a two-atom dark state in the presence of strong blockade and no dynamical phase is accumulated\cite{Moller2008}. However, such an approach is not useful when $\ket{p}$ is subject to radiative decay\cite{Petrosyan2013}. 
Instead, we use 
$\Delta_1\ne 0$ so there is minimal excitation of state $\ket{p}$. The dynamics do not follow a dark state, and the $2\pi$ rotation $\ket{11}\rightarrow \frac{\ket{1r}+\ket{r1}}{\sqrt2}\rightarrow\ket{11}$ gives a dynamical phase of $\phi_2=\pi$. 

Using the parameters of Fig.~\ref{fig.stirap1} and the same steps as for the ARP protocol we prepare the Bell state $\ket{B'}=\frac{\ket{01}+\ket{10}}{\sqrt2}$. This is different than the state prepared with ARP pulses due to the different choices of $\phi_1, \phi_2$. 
 For the parameters of Fig.~\ref{fig.stirap1} we find that the state  $\ket{B'}$ is created with fidelity 
${\mathcal F} = 0.976, .978, .979$ at ${\sf B}/2\pi=500, 1500, 3000~\rm MHz$.

The fidelity of the STIRAP gate is substantially lower than that achieved with ARP pulses. The reason is that the parameters used are not sufficiently adiabatic giving imperfect following of the dark state. 
Calculations show that $\phi_1= -18.~\rm  deg.$ and $\phi_2= -171.5~\rm  deg.$.  So there is a phase error of 18 deg. for one atom excited and 9.5 deg. for two atoms.  Although tests with slower, more adiabatic pulses, with spontaneous emission turned off result in Bell states with arbitrarily high fidelity the challenge is to design pulse shapes that are both adiabatic and sufficiently fast to prevent spontaneous emission errors.  One approach may be to compensate the imperfect dynamical phase with a geometrical phase by adjusting the relative phase of $\Omega_1$ and $\Omega_2$\cite{Moller2007}. 

%%%%%%%%%%%%%%%%FIGURE%%%%%%%%%%%%%%%%
\begin{figure}[!t]
\includegraphics[width=8.2cm]{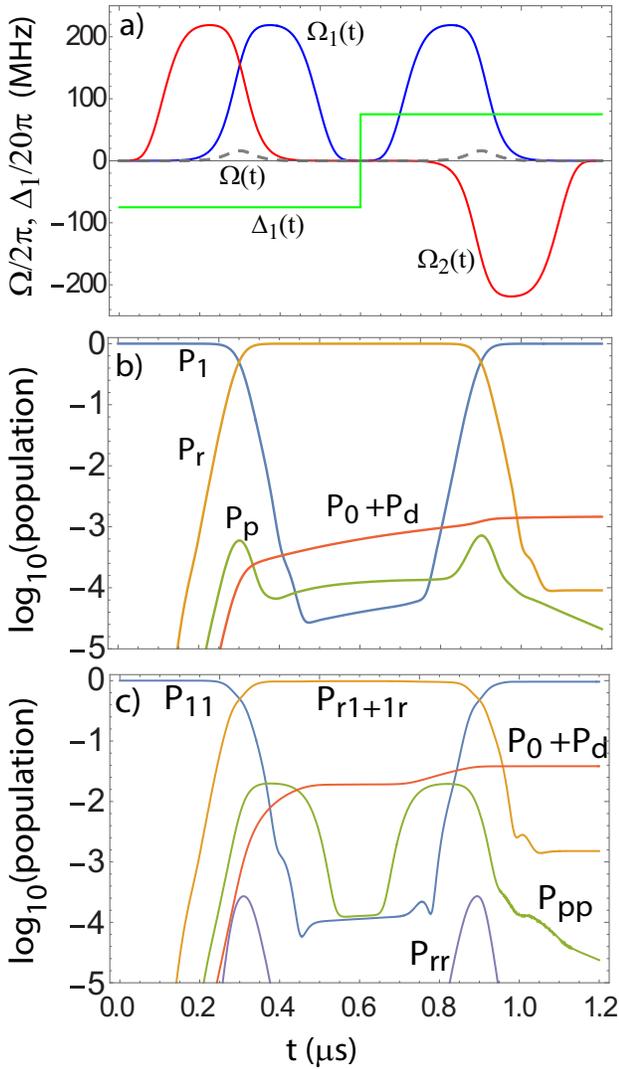}
\caption{(color online)
$C_Z$ gate with STIRAP pulses and total time of $T=1.2~\mu\rm s$. a) Pulse shapes as defined in Eqs.~(\ref{eq.stirap2}) with 
$\Omega_0/2\pi=220~\rm MHz$,
$\Delta_1/2\pi=750~ {\rm sign}(t-T/2)~\rm MHz$,
$t_1=0.3~\mu\rm s,$ 
$t_2=0.9~\mu\rm s,$ and
$\tau=0.1~\mu\rm s$. All other parameters the same as in Fig.~\ref{fig.stirap1}.
  b) Populations of one atom states for the initial state $\ket{10}$ and c) populations of two atoms states for the initial state $\ket{11}$. Leakage population curves as defined in Fig.~\ref{fig.stirap1}.   
}
\label{fig.stirap2}
\end{figure}

Alternatively we may use optimized analytical pulse shapes~\cite{Vasilev2009} to improve the entanglement fidelity as shown in Fig.~\ref{fig.stirap2}. 
This design uses a double STIRAP sequence with switching of the sign of the detuning and the phase of the $\Omega_2$ pulse halfway through the gate.
This is a modification of the scheme considered earlier in  
\cite{Beterov2013a,Beterov2016b}. To achieve high fidelity we use the optimized pulse shapes~\cite{Vasilev2009}
\begin{subequations}
\begin{eqnarray}
\Omega_1(t)&=&\Omega_0 F(t-t_1)\sin\left[\frac{\pi}{2}f(t-t_1) \right]\nonumber\\
&&+ \Omega_0 F(t-t_2)\cos\left[\frac{\pi}{2}f(t-t_2) \right],\\
\Omega_2(t)&=&\Omega_0 F(t-t_1)\cos\left[\frac{\pi}{2}f(t-t_1) \right]\nonumber\\
&&- \Omega_0 F(t-t_2)\sin\left[\frac{\pi}{2}f(t-t_2) \right],
\end{eqnarray}
\label{eq.stirap2} 
\end{subequations}
with $F(t)=e^{-(t/2\tau)^{6}}$ and $f(t)=\left(1+e^{-4t/\tau} \right)^{-1}.$ The times $t_1, t_2$ correspond to the centers of the $\Omega_1$ pulses shown in Fig.~\ref{fig.stirap2}a). Note that in addition to the sign of the intermediate state detuning changing in the middle of the gate, the phase of $\Omega_2$ changes by $\pi$ in the second half of the gate. 
In the ideal case of no spontaneous emission and infinite blockade these pulses result in the state evolution 
$\ket{00}\rightarrow \ket{00}, 
\ket{01}\rightarrow -\ket{01}, 
\ket{10}\rightarrow -\ket{10}, 
\ket{11}\rightarrow -\ket{11}$
 which implements a controlled phase gate. Numerical simulations of the dynamical evolution with finite blockade and Rydberg lifetime are shown in Fig.~\ref{fig.stirap2}.
The predicted Bell state fidelity is 
${\mathcal F}=0.990, 0.991$ for ${\sf B}/2\pi=500, 1500~\rm MHz$. The infidelity is reduced by about a factor of 2 relative to the pulse scheme of Fig.~\ref{fig.stirap1}
although the dominant error source is still  scattering from the intermediate 
$\ket{p}$ state leading to growth of population in the uncoupled ground state $\ket{d}$. This error can be reduced using larger $\Delta_1$ and the Bell state fidelity increased to 0.996, but only with unrealistically high Rabi frequencies. 

\subsection{$C_Z$ gate with numerically optimized pulses}
\label{sec.STIRAP2}

%%%%%%%%%%%%%%%%FIGURE%%%%%%%%%%%%%%%%
\begin{figure}[!t]
\includegraphics[width=8.5cm]{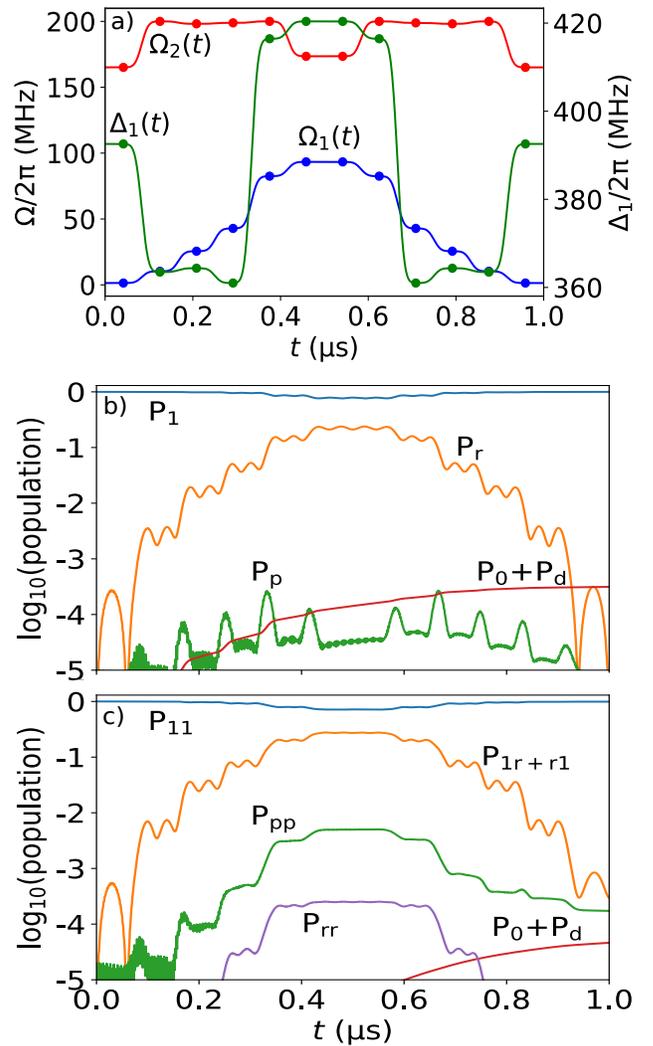}
\caption{(color online)
$C_Z$ gate with globally optimized 12 segment pulses for $T=1.0~\mu\rm s$. 
a) Shapes of $\Omega_1(t),~ \Omega_2(t)$ and $\Delta_1(t)$. The maximum slew rate of the pulses was limited to 1 GHz$/\mu\rm s$.  
b) Population of the states $\ket{10}$, $\ket{p0}$, $\ket{r0}$,  and $\ket{00}+\ket{d0}$ states for the initial state  $\ket{10}$. 
c) Population of the states $\ket{11}$, $\ket{pp}$, $\ket{1r}+\ket{r1}$, $\ket{rr}$ and $\ket{0}+\ket{d}$ states for the initial state  $\ket{11}$.
The blockade strength was ${\sf B}/2\pi=500~\rm MHz$ and all other parameters were the same as in Fig.~\ref{fig.stirap2}.  }
\label{fig.stirap3}
\end{figure}

A higher fidelity version of the two-photon  adiabatic gate can be developed by designing the time-dependent intensity and detuning of the control lasers such that $\phi_1=0$ and $\phi_2=\pi$. We use the same two-atom Hamiltonian as described in Sec.~\ref{sec.STIRAP} to model the gate procedure. Using pulse profiles that don't require high accuracy population transfer between ground and Rydberg states we retain the adiabatic character of the gate while achieving higher Bell-state fidelity relative to the use of STIRAP pulses.

We employ global optimization algorithms to search for pulse profiles that deliver feasible pulse sequences that improve the gate fidelity.
In this approach we divide $f(t)$ $(\Omega_1(t),$ $\Omega_2(t),$ or  $\Delta_1(t)$)  into $2N$ equal length segments that are piece-wise continuous and parameterized by the magnitude of each segment $f_i(t)$, with the segments connected  by error functions according to 
\begin{equation}
f(t)=\frac{f_i+f_{i+1}}{2}+\frac{f_{i+1}-f_i}{2}\text{erf}\left[\frac{5}{\Delta t}\left(t-\frac{t_i+t_{i+1}}{2}\right)\right],\nonumber
\end{equation}
where $t_i\le t \le t_{i+1}$ for the $i^\text{th}$ segment, each of which has length $\Delta t$~\cite{Zahedinejad2016}. The use of error functions
as building blocks facilitates constructing smooth pulse shapes that 
respect constraints imposed by available devices for optical 
modulation. Using pulses that are symmetric about the central time of the gate, the optimization problem is reduced to finding $3N$ independent variables (for $\Omega_1(t), \Omega_2(t), \Delta_1(t)$) that optimize the entanglement fidelity $\mathcal F$ while satisfying a constraint on maximum modulation rate. 

As the above optimization problem is highly
constrained with a non-convex objective function $\mathcal F$,
we employ an off-the-shelf global optimization algorithm to solve the problem. We
use a parallelized version of differential evolution~\cite{Storn1997} motivated by previous work on quantum control problems~\cite{Zahedinejad2014}. Optimized pulse shapes for $N=6$ are shown in Fig.~\ref{fig.stirap3}a) and the coefficients for each segment are listed in Table \ref{tab.segments}. These pulses give ${\mathcal F}=0.997$ for a gate time of 1 $\mu\rm s$, which is significantly improved compared to STIRAP with analytic pulse shapes, and is slightly better than the fidelity of the ARP gate at the same blockade strength.  

\begin{table}[!t]
\caption{Segment coefficients (in MHz) for the  pulses in Fig.~\ref{fig.stirap3}a). \label{tab.segments}}
\begin{center} 
\begin{tabular}{ |c|c|c|c|c|c|c| }
 \hline
  & \multicolumn{6}{ | c | }{Segment number} \\
\hline
Functions  & 1 \& 12 & 2 \& 11 & 3 \& 10 & 4 \& 9 & 5 \& 8 & 6 \& 7 \\
 \hline
 $\Omega_1(t)/2\pi$ & 1.38 & 10.30 & 25.54 & 42.85 & 82.50 & 93.35 \\ 
 $\Omega_2(t)/2\pi$ & 165.09 & 199.99 & 198.14 & 198.87 & 200.00 & 173.48\\ 
 $\Delta_1(t)/2\pi$ & 392.57 &  363.48 &  364.36 &  360.99 &  416.45 & 420.39\\ 
 \hline
\end{tabular}
\end{center}
\end{table}

The optimization done using the Hamiltonian in Eq.~(\ref{eq.H2}) might rely on non-adiabatic evolution, nevertheless we observe in Fig.~\ref{fig.stirap3} that the evolution of the system lies mainly on the states $\ket{11}$ and $\ket{1r}+\ket{r1}$, with negligible population of the $\ket{0}$ and  $\ket{d}$ states.  In this regard, the overlap of the system state with the instantaneous dark state is always $\sim 1$ for the case when the initial state is $\ket{10}$ or $\ket{01}$. This fact is also evident from evaluating  the mixing angle 
defined by $\operatorname{tan}^{-1}\left[\frac{\Omega_1(t)}{\Omega_2(t)}\right]$
whose maximum rate of change remains much smaller than the Rabi rate at each instant of time.
In addition, the  mixing angle changes smoothly from 0 to 0.5 rad till the middle of the gate protocol leading to incomplete population inversion, instead of 0 to $\pi/2$ rad for the usual STIRAP protocol. We therefore refer to this pulse  as STIRAP inspired. 
 The result of these simulations reveal an accumulated dynamical phase $\phi_1$  and $\phi_2$ of -1.3 deg. and 178.8 deg., respectively. This optimized pulse sequence demonstrates how the errors in $\phi_1$ and $\phi_2$ are much smaller than in the conventional case analyzed in Sec.~\ref{sec.STIRAP1} leading to a high Bell-state fidelity.

\begin{figure}[!t]
    \centering
    \includegraphics[width=8.5cm]{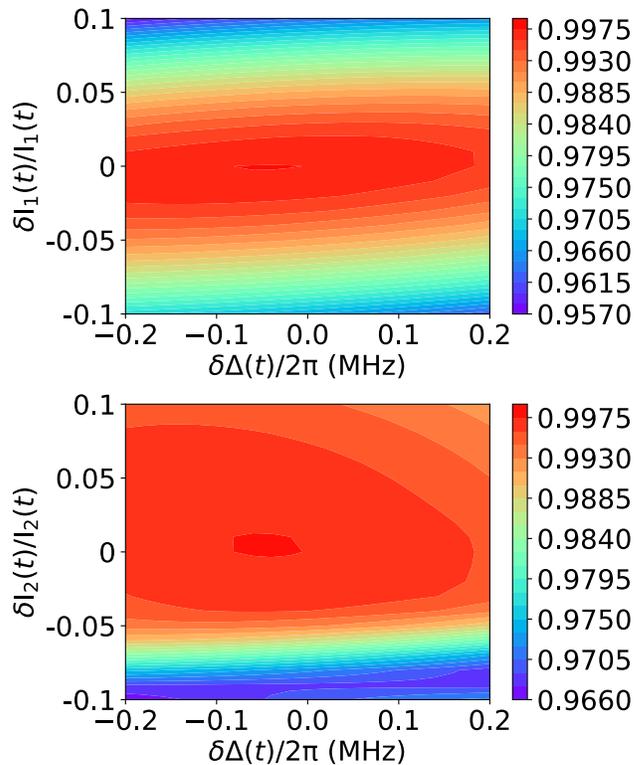}
    \caption{(color online) Bell fidelity sensitivity to variations in pulse amplitude and detuning with parameters from  Fig.~\ref{fig.stirap3}a). The contour plots show $\mathcal{F}$ as a function of fractional change in the intensity of the lasers and changes in the two-photon detuning.}
    \label{fig.optimizedSTIRAPsensitivity}
\end{figure}

The robustness of the optimized pulse sequence is shown in  Fig.~\ref{fig.optimizedSTIRAPsensitivity} for the same parameters as in Fig.~\ref{fig.stirap3}. We consider a laser intensity fluctuation of $\pm 10~\%$ and a laser frequency change of $\pm200$ kHz. At the limits of the frequency variation the fidelity is reduced by about 0.005, and at the limits of the intensity variation the fidelity is reduced by at most 0.04.  Comparing to the sensitivity of the standard gate with constant amplitude pulses and $1.0~\mu\rm s$ duration in  the Appendix Fig. \ref{fig.appendix}b) we see that there is about $6\times$ less  sensitivity to detuning and about twice higher  sensitivity to intensity noise.

\section{Discussion}
\label{sec.discussion}

We have presented adiabatic  pulses using ARP, STIRAP, and STIRAP-inspired optimized pulses that lead to  high Rydberg gate fidelities with reduced sensitivity to variations of the amplitude and frequency of the laser pulses. For the ARP protocol
of Fig.~\ref{fig.arp1} we obtain Bell fidelity  ${\mathcal F}=0.9994$ at a blockade strength of ${\sf B}/2\pi=3~\rm GHz$ and  gate duration of 0.54 $\mu\rm s$. We show in Fig.~\ref{fig.arp_robustness} that the sensitivity to parameter variations is reduced by an order of magnitude compared to the standard protocol with constant amplitude pulses.

For the STIRAP pulses of Figs.~\ref{fig.stirap1} and \ref{fig.stirap2} we obtain a lower fidelity of  ${\mathcal F}\simeq 0.98, 0.99$ with experimentally realistic parameters. The dominant  errors are scattering from the intermediate state used for Rydberg excitation and imperfect following of the adiabatic dark state. These errors can be reduced but only at the cost of very high laser powers. 
In Fig.~\ref{fig.stirap3} we present a STIRAP inspired globally optimized pulse protocol that reaches  ${\mathcal F}=0.997$ at a blockade strength of ${\sf B}/2\pi=500~\rm MHz$ and  gate duration of 1.0 $\mu\rm s$.  This protocol uses time dependent amplitude and detuning that are inspired by, yet distinct from the usual STIRAP sequence. Nevertheless we verify that the dynamics follow a dark state evolution.  The protocol has reduced sensitivity to detuning errors, as shown in Fig. \ref{fig.optimizedSTIRAPsensitivity}, but higher sensitivity to intensity variations compared to the standard protocol with constant amplitude pulses.

All the pulses analyzed here are applied simultaneously to both atoms, which removes the need for high speed switching of lasers between different 
spatial locations. This may be advantageous for gate operations in large qubit arrays as in \cite{Graham2019}, as well as for simultaneous operation of multiple gates on atom pairs that are far enough apart that the Rydberg interaction does not cause crosstalk.  The results account fully for spontaneous emission from all participating excited atomic states in a room temperature environment. The reported gate fidelities are defined as the fidelity of a Bell state prepared by the gate assuming there are no other control errors,  zero excess laser noise, and no errors due to atomic motion or crosstalk in a multi-qubit array. For a detailed exposition of technical error sources we refer to the supplemental material in \cite{Graham2019}. Compared to Rydberg gates using constant amplitude, or other non-adiabatic pulses, we show that ARP and STIRAP inspired protocols  provide improved robustness in the presence of Doppler shifts at finite atomic temperature. 

These results, while promising, should not be considered as an ultimate limit on the Rydberg gate fidelity with simultaneous addressing of both atoms. Due to the fact that the Rydberg states have a finite lifetime the gate error is lower bounded by the integrated population of the Rydberg states during the gate. In order to prepare an entangled state this integrated population cannot be arbitrarily small, as has been pointed out in \cite{Wesenberg2007}. Therefore optimization of the gate fidelity is not just a matter of running the gate arbitrarily fast which would not lead to entanglement, but of finding high fidelity and robust pulse sequences at finite speed.  The space of possible pulse designs is large, and further exploration allowing for a wider range of pulse shapes with more degrees of freedom to optimize over is likely to lead to even higher fidelity limits.

%%%%%%%%%%%%%%%%%%%%%%%%%%%%%%%%
\acknowledgments
MS was supported by the ARL-CDQI Center for Distributed Quantum Information,  
NSF PHY-1720220, DOE award DE-SC0019465, and ColdQuanta, Inc.
MS is grateful to David Petrosyan for useful comments on the manuscript. 
IIB was supported by the Russian Foundation for Basic Research under Grant No. 17- 02-00987
(for numeric simulations),
and by the Russian Science Foundation under Grant No. 18-12-00313. 
AD would like to thank Mitacs Inc., the Canadian Queen Elizabeth II Diamond Jubilee Scholarships program (QES) and ARL grant W911NF-18-1-0203 for funding support
and Compute Canada Calcul Canada for computational support. BCS appreciates financial support from NSERC and MIF from Government of Alberta, Canada. AD, EJP and BCS thank S. S. Vedaie for useful discussions on the differential evolution algorithm.

\input{CZgate_resubmit.bbl}

%\bibliography{d:/users/mark/pubs/biblio/saffman_refs,d:/users/mark/pubs/biblio/rydberg,d:/users/mark/pubs/biblio/qc_refs,d:/users/mark/pubs/biblio/atomic,d:/users/mark/pubs/biblio/holmium_v1}

\appendix

\section{Sensitivity analysis of standard Rydberg gate protocol}

We wish  to compare the sensitivity of the adiabatic protocols to the standard protocol
that uses constant amplitude pulses. To do so we derive an analytic expression for the gate error due to variations in detuning $\Delta$ and optical intensity $\delta I$ that change the pulse areas away from their ideal values. In order to simplify the analysis we assume the ideal limit of $\omega_q, {\sf B}\gg \Omega\gg 1/\tau$ and neglect errors due to finite blockade strength and finite Rydberg state lifetime. This allows us to extract the sensitivity to variations in  pulse parameters. A detailed analysis of the standard protocol including finite blockade and Rydberg lifetime errors was given in \cite{XZhang2012}.

In this limit the standard protocol~\cite{Jaksch2000} of $\pi$ pulse on control atom, $2\pi$ pulse on target atom, $\pi$ pulse on control atom can be represented with the evolution operator 
\begin{eqnarray}
U_{\rm Z, ideal}&=&\left[R(\pi)\otimes (P_{\ket{0}}+P_{\ket{1}})+I\otimes P_{\ket{r}}\right] \nonumber\\
&\times &\left[(P_{\ket{0}}+P_{\ket{1}})\otimes R(2\pi)+P_{\ket{r}}\otimes I\right]\nonumber\\
&\times&\left[R(\pi)\otimes (P_{\ket{0}}+P_{\ket{1}})+I\otimes P_{\ket{r}}\right].
\label{eq.Uideal}
\end{eqnarray}
Here the $R(\theta)$ operator is a rotation with area $\theta$ between states $\ket{1}\leftrightarrow \ket{r}$ and $P_{\ket{j}}$ is a projector onto state $\ket{j}$. The operator $U_{\rm ideal}$ assumes perfect blockade so when the other atom is in state $\ket{r}$ the pulse is  
completely blocked. 

A perfect Bell state is prepared with the sequence
\begin{equation}
\ket{\rm Bell}
=
(I\otimes H)U_{Z, \rm ideal}(I\otimes H)(H\otimes I)\ket{00},
\label{eq.Bellideal}
\end{equation}
where $H$ is the Hadamard gate acting on the $\ket{0},\ket{1}$ qubit states. 
Starting with the input state $\ket{00}$ this sequence generates $$
\ket{{\rm Bell}}=\frac{\ket{01}-\ket{10}}{\sqrt2}.
$$

Variations in the detuning $\Delta$ or optical intensity $\delta I$ lead to errors in the $\pi$ and $2\pi$ Rydberg pulses. These errors  are primarily  due to\cite{Graham2019} laser frequency instability, Doppler shifts from atomic motion, external fields shifting the Rydberg energy,
variations in the optical intensity from laser noise, atomic position variations relative to the control beams, or optical pointing fluctuations. 
The rotation operator accounting for these errors can be expressed in the basis $\{\ket{0},\ket{1},\ket{r}\}$ as 
\begin{widetext}
\begin{equation}
R(t,\Omega_0,\Delta,\delta I)=
\begin{pmatrix}
1 & 0 & 0\\
0&e^{\imath \Delta t/2}\left[\cos(\Omega' t/2) - i \frac{\Delta}{\Omega'}\sin(\Omega' t/2) \right]&ie^{\imath \Delta t/2}\frac{\Omega_0^*\sqrt{1+\delta I}}{\Omega'}\sin(\Omega' t/2)\\
0&ie^{\imath \Delta t/2}\frac{\Omega_0\sqrt{1+\delta I}}{\Omega'}\sin(\Omega' t/2)&e^{\imath \Delta t/2}\left[\cos(\Omega' t/2) + i \frac{\Delta}{\Omega'}\sin(\Omega' t/2) \right]
\end{pmatrix}
\label{eq.Rerror}
\end{equation}
\end{widetext}
with $\Omega'=\sqrt{|\Omega_0|^2(1+\delta I)+\Delta^2}$. The rotation operator has been expressed in a frame rotating at the frequency $\omega$ of the applied optical field and $\Delta=\omega-\omega_{r1}$ is the detuning from the atomic transition.  Non-ideal $\pi$ and $2\pi$ pulses are expressed as 
$R(\pi/|\Omega_0|,\Omega_0,\Delta,\delta I)$
and 
$R(2\pi/|\Omega_0|,\Omega_0,\Delta,\delta I)$.
The operator generating a  Bell state accounting for pulse errors $\ket{{\rm Bell}'}$ is then found from Eq. (\ref{eq.Bellideal}) with $R(\pi), R(2\pi)$ replaced by the corresponding expressions from (\ref{eq.Rerror}).  

\begin{figure}[!t]
    \centering
    \includegraphics[width=.48\textwidth]{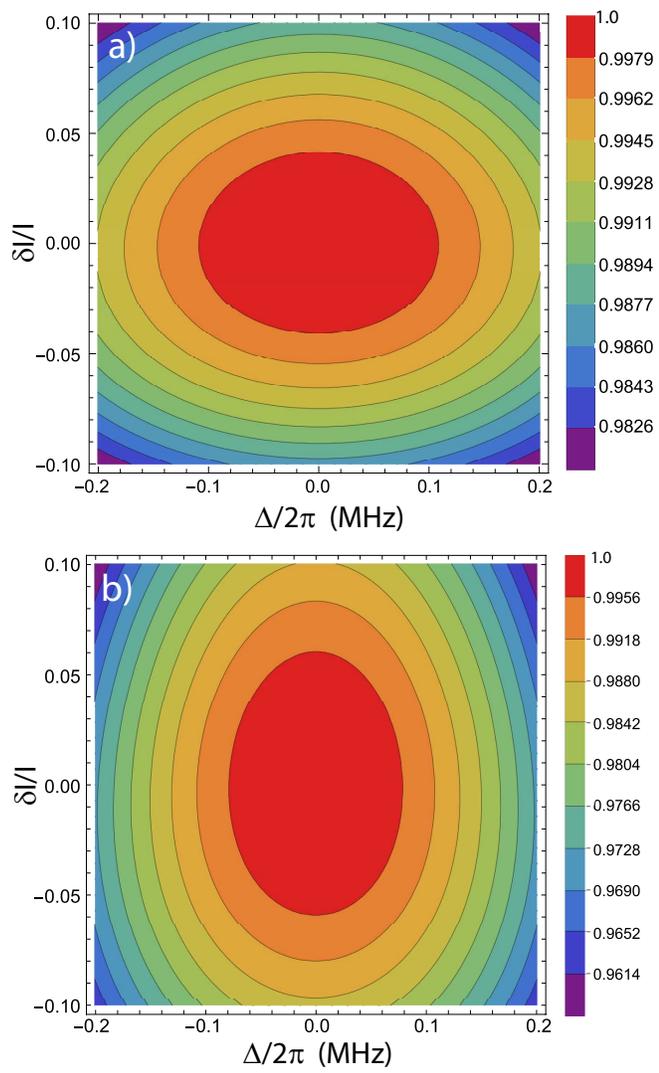}

    \caption{\label{fig.appendix}Fidelity of the idealized Rydberg gate with constant pulse amplitude for variations in detuning and intensity and a gate duration of a) $0.5~\mu\rm s$, b) $1.0~\mu\rm s$.}
\end{figure}

Following this procedure we generate an analytical but lengthy expression for 
$\ket{{\rm Bell}'}$ and quantify the sensitivity to pulse errors using the pure state fidelity expression 
${\mathcal F}=|\bra{{\rm Bell}'}{\rm Bell}\rangle|^2.$
For $\Delta=\delta I = 0$ the fidelity is ${\mathcal F}=1$.
In order to make a direct sensitivity comparison 
with the results for adiabatic protocols  we have set $\Omega_0/2\pi=4(2) ~\rm MHz $ so the gate time is $0.5(1.0)~\mu\rm s$ which is the same as that used in Figs. \ref{fig.arp1}, \ref{fig.stirap3}. The results are shown in Fig. \ref{fig.appendix}. Compared to Figs. \ref{fig.arp_robustness}, \ref{fig.optimizedSTIRAPsensitivity} 
we see that the sensitivity to both detuning and intensity errors is much higher than for the ARP protocol and the sensitivity to the detuning error is much higher than for the STIRAP inspired protocol.

\end{document}

%% file: CZgate_resubmit.bbl

%% file: CZgate_resubmit.bbl
\begin{thebibliography}{41}
\bibitem [{201()}]{2019stirapnote}%
  \BibitemOpen
  \href@noop {} {}\bibinfo {note} {With two-photon excitation the Rydberg state
  is either $ns$ or $nd$. To facilitate comparison with the ARP gate we have
  used a Rydberg lifetime of 540. $\mu\rm s$ corresponding to the Cs
  107$p_{3/2}$ state as in Fig. \ref{fig.arp2}. A similar lifetime occurs for
  Cs $126s_{1/2}$ or $132d_{5/2}$ states.}\BibitemShut {Stop}%
\end{thebibliography}
